\documentclass[12pt]{article}
\usepackage{a4wide}
\usepackage{amssymb, amsmath}
\usepackage[normalem]{ulem}
\usepackage{graphicx}
\usepackage{cite}
\usepackage[utf8]{inputenc}
\usepackage[colorlinks]{hyperref}
\usepackage[dvipsnames]{xcolor}
\usepackage{subcaption}
\usepackage[justification   = raggedright,
	singlelinecheck = false]{caption}

\newcommand{\ev}{{\cal E}}
\newcommand{\s}{{\cal S}}
\newcommand{\h}{H_{\theta}}

\newcommand{\cpp}[2][-]{\hat{c}_{#1 \bf #2}^{\dagger}} 
\newcommand{\cm}[2][]{\hat{c}_{#1 \bf #2}} 

\newcommand{\vt}[1]{{\small \bf #1}} 
\newcommand{\dpr}[2]{{\bf #1} \cdot {\bf #2}} 
\newcommand{\Mp}{M_{\rm Pl}} 
\newcommand{\ket}[1]{\left| #1 \right\rangle} 
\newcommand{\bra}[1]{\left\langle #1 \right|} 
\newcommand{\enm}[3][]{\hat{\chi}_{#1\vt{#2}}^{_{\ev}}(#3)} 
\newcommand{\ssm}[3][]{\hat{\chi}_{#1\vt{#2}}^{_{\s}}(#3)} 
\newcommand{\Tr}{{\rm Tr}}

\begin{document}
{\renewcommand{\thefootnote}{\fnsymbol{footnote}}
		
\begin{center}
{\LARGE Universal signature of quantum entanglement\\ across cosmological distances} 
\vspace{1.5em}

Suddhasattwa Brahma$^{1}$\footnote{e-mail address: {\tt suddhasattwa.brahma@gmail.com}}, 
Arjun Berera$^{2}$\footnote{e-mail address: {\tt ab@ed.ac.uk}}, and 
Jaime Calder\'on-Figueroa$^{2}$\footnote{e-mail address: {\tt jaime.calderon@ed.ac.uk}}
\\
\vspace{1.5em}
$^1$  Department of Physics, McGill University, Montr\'eal, QC H3A 2T8, Canada\\

\vspace{0.5em}
$^2$  School of Physics and Astronomy, University of Edinburgh\\
Edinburgh, EH9 3FD, United Kingdom\\
\vspace{1.5em}
\end{center}
}
	
\setcounter{footnote}{0}

\newcommand{\bea}{\begin{eqnarray}}
\newcommand{\eea}{\end{eqnarray}}
\renewcommand{\d}{{\mathrm{d}}}
\renewcommand{\[}{\left[}
\renewcommand{\]}{\right]}
\renewcommand{\(}{\left(}
\renewcommand{\)}{\right)}
\newcommand{\nn}{\nonumber}
\newcommand{\Mpl}{M_{\textrm{Pl}}}
\def\H{\mathrm{H}}
\def\V{\mathrm{V}}
\def\e{\mathrm{e}}
\def\be{\begin{equation}}
\def\ee{\end{equation}}

\def\al{\alpha}
\def\bet{\beta}
\def\gam{\gamma}
\def\om{\omega}
\def\Om{\Omega}
\def\sig{\sigma}
\def\Lam{\Lambda}
\def\lam{\lambda}
\def\ep{\epsilon}
\def\ups{\upsilon}
\def\vep{\varepsilon}
\def\S{\mathcal{S}}
\def\doi{http://doi.org}
\def\arxiv{http://arxiv.org/abs}
\def\d{\mathrm{d}}
\def\g{\mathrm{g}}
\def\m{\mathrm{m}}
\def\r{\mathrm{r}}

\begin{abstract}
	\noindent Although the paradigm of inflation has been extensively studied to demonstrate how macroscopic inhomogeneities in our universe originate from quantum fluctuations, most of the established literature ignores the crucial role that \textit{entanglement} between the modes of the fluctuating field plays in its observable predictions. In this paper, we import techniques from quantum information theory to reveal hitherto undiscovered predictions for inflation which, in turn, signals how quantum entanglement across cosmological scales can affect large scale structure. Our key insight is that observable long-wavelength modes must be part of an \textit{open quantum system}, so that the quantum fluctuations can decohere in the presence of an environment of short-wavelength modes. By assuming the simplest model of single-field inflation, and considering the leading order interaction term from the gravitational action, we derive a \textit{universal lower bound} on the observable effect of such inescapable entanglement.
\end{abstract}

\noindent Non-local behaviour, manifested through quantum entanglement, is a characteristic trait of quantum theory. An implication of this is that physics on different scales are able to influence each other. Conversely, the Wilsonian prescription tells us that it is possible to derive effective field theories (EFTs) which can describe physics below a particular energy scale by integrating out high-energy degrees of freedom (dofs). Therefore, it is important to understand what are the conditions under which a quantum system allows for a derivative expansion, or some analogous controlled approximation, for having the usual EFT description.

An early epoch of accelerated expansion \cite{starobinsky1980new, Fang:1980wi, guth1981inflationary, Sato:1980yn, linde1982new, albrecht1982cosmology} -- inflation -- presents a rare example of interplay between physics on microscopic and macroscopic scales since it explains the rich structure of late-time inhomogeneities, observed in the distributions of galaxies and the temperature variations in the cosmic microwave background (CMB), as being sourced by minuscule quantum mechanical vacuum fluctuations \cite{Mukhanov:1982nu}. Primordial quantum fluctuations, whose physical wavelengths are stretched by the background expansion, become larger than the comoving Hubble radius at some point in the remote past and then re-enter as classical perturbations which we observe today. However, since the modes on super-Hubble scales alone form our system,  we need to figure out which is the best EFT that describes the dynamics of these long-wavelength dofs. 

The crucial thing to note in this case is that the standard Wilsonian effective action does not exist since the sub-Hubble modes, which are integrated out, are not excluded by any conservation law \cite{Burgess:2014eoa,Burgess:2015ajz}. This is contrary to traditional EFTs in which energy conservation ensures that high-energy dofs cannot be part of the system if they were not initially present, and the entire dynamics of the low-energy system can be described by an effective Lagrangian consisting only of the light dofs. In the cosmological system, the long wavelength modes interact and exchange energy with the sub-Hubble ones and therefore, the time evolution of the system under consideration cannot be described by unitary evolution of some low-energy Hamiltonian.  In fact, these long wavelength dofs need to be treated as an open quantum system rather than an isolated one \cite{Shandera:2017qkg}, which is coupled to an environment of unobservable short wavelength fluctuations, thereby allowing for a pure state to evolve into a mixed one \cite{Burgess:2006jn}. The non-Hamiltonian evolution underlying this process is the main physical insight which allows physics on the shortest scales to affect structure formation on the largest scales of our universe, going against intuition derived from Wilsonian EFT since we need to study an \textit{out-of-equilibrium} system in this case. The key question, which we address in this article, is what are the observable consequences of this primordial entanglement between the long and short wavelength modes?

The simplest model of inflation consists of a (minimally-coupled) single scalar field, with a postulated form for its potential to allow for an accelerated expansion and then a graceful exit from it. Gravitational nonlinearities, arising from the Einstein-Hilbert action \cite{Maldacena:2002vr}, lead to coupling between the momentum modes (in Fourier space) of cosmological perturbations. We consider the coupling between the band of super-Hubble modes and unobservable short-wavelength ones to be given by gravity and, is thus, \textit{inescapable}. From more general considerations, one expects cosmological perturbations to be be gravitationally coupled to other dofs present in the universe, and this shall only result in a larger amount of entanglement, and a subsequent magnification of our result. Consequently, we present a \textit{universal lower bound} on the observable consequences of primordial quantum entanglement  -- universal since this is due to interactions arising from pure gravity (and must, therefore, be present in any theory which has general relativity as its low energy limit) and it is valid for \textit{any} model of inflation, since additional fields will only lead to a bigger effect (see {\it e.g.} \cite{Boyanovsky:2015tba, Boyanovsky:2018soy,Boyanovsky:2018fxl}).

The novelty of our work is twofold: Firstly, we consider corrections to the primordial spectra of observable modes coming from quantum entanglement due to considering an \textit{open EFT} and secondly, we consider interactions between long and short dofs (where the Hubble horizon of the quasi-de Sitter (dS) background sets the reference scale) of the inflaton fluctuations themselves (as opposed to considering additional postulated fields) due to gravitational nonlinearities. Even on a conceptual level, the quantum-to-classical transition of cosmological perturbations necessarily rely on treating them as part of an open system, thereby allowing for decoherence. This is why we need to use techniques for open EFTs in which one studies the (non-unitary) evolution of the reduced density matrix obtained by tracing over the unobservable dofs. These methods are prevalent in other branches of physics, and we adapt them here for cosmology, in the hope of measuring the degree of entanglement, between our super-Hubble system modes with its environment (see \cite{Sharman:2007gi,Brahma:2020zpk} for entanglement entropy calculations in this setup), through observable predictions. Therefore, we strengthen the bonds between cosmology and quantum information theory by uncovering new observable effects in the former, which have remained out of bounds of closed quantum system treatments with Wilsonian EFTs, by importing mathematical tools from the latter. 

Our main findings are that there are non-negligible corrections to the power spectrum arising from this primordial entanglement due to UV-IR mode-mixing, which are detectable from future observations. Secondly, in order for the perturbative description to be valid, inflation cannot be semi-infinite in the past since the small corrections from primordial entanglement build up with time. A by-product of our work is that it serves as an indirect test for cubic non-Gaussianities in vanilla slow-roll models of inflation, which cannot otherwise be detected directly from any observation \cite{Chen:2010xka,Pajer:2013ana}. Since these leading order nonlinearities necessarily act as an interaction term responsible for mode-couplings between the system and environment dofs, verification of our prediction shall indirectly validate their existence. (More complicated models of inflation, which allow for detectable amounts of non-Gaussianity, will only enhance this effect.) Finally, since entanglement is a purely quantum phenomenon, our results present a smoking gun for the quantum origin of inflation. In principle, present observations of macroscopic inhomogeneities can be modelled by a classical probability distribution and an unequivocal test of their quantum origin (such as cosmological Bell inequalities \cite{Campo:2005sy, Campo:2005sv, Maldacena:2015bha,Martin:2016tbd} or specific features in the non-Gaussianity functions \cite{Green:2020whw}) are typically too small for realistic models of inflation \cite{Martin:2017zxs,Morse:2020mdc}.  A null detection of our predictions, from future observations, will definitively rule out a quantum origin for inflation since we give a universal lower bound of a verifiable effect which must be present for quantum fluctuations resulting from a single scalar field dynamics. Our methods, however, are completely general and lay the foundation for calculating the same effect for any theory of the early-universe, \textit{e.g. Ekpyrosis} \cite{Khoury:2001bz,Khoury:2001wf}, which sources inhomogeneities from quantum fluctuations.

The quasi-dS space can be described by the homogeneous metric: $ds^2=-dt^2 +a^2(t) d{\bf x}^2=-a^2(\tau) \left(d\tau^2-d{\bf x}^2\right)$ where the scale factor $a$, Hubble parameter $H=\dot{a}/a$ and the slow-roll parameters $\epsilon:= -\dot{H}/H^2,\;\eta:=\dot{\epsilon}/(H\epsilon)$ describe evolution of the background. Overdots and primes denote derivatives with respect to cosmic time `$t$' and conformal time `$\tau$', respectively, which are related by $a d\tau = dt$. Throughout, we shall assume that $H$ is almost constant and the slow-roll parameters remain small during the entire evolution. We denote the comoving curvature perturbation as $\zeta$, and the canonical Mukhanov-Sasaki variable as $\chi = z(\tau) \zeta$, where $z^2 = 2 \epsilon a^2 \Mp^2$, where $\Mp$ is the reduced Planck mass. The underlying idea is that this comoving curvature perturbation, which is a combination of the quantum fluctuations of matter and the linearized gravitational field, can be described as a set of harmonic oscillators to leading order (in Fourier space), accentuated by interaction terms which result due to the non-linearity of the gravitational action. 

The quadratic action for the canonical variable reads $\mathcal{L}^{(2)} = \int d^3 x\ \left[(\chi')^2 - (\partial \chi)^2 + \frac{z''}{z} \chi^2 \right]$, the solutions for which describe the dynamics of the mode functions to be used below. This is nothing but the Lagrangian for a harmonic oscillator with a time-dependent mass term. Going to Fourier space and introducing the usual ladder operators, the quadratic Hamiltonian
\begin{eqnarray}\label{eq:Hml}
	\hat{H}^{(2)} &=& \frac{1}{2} \int \frac{d^3k}{(2\pi)^3} \left(k \left[\hat{c}_{\bf k} \hat{c}^\dagger_{\bf k} +\hat{c}_{\bf -k} \hat{c}^\dagger_{\bf -k}\right] - i \frac{z'}{z} \left[\hat{c}_{\bf k} \hat{c}_{\bf -k} - \hat{c}^\dagger_{\bf k} \hat{c}^\dagger_{\bf -k}\right]\right)\,
\end{eqnarray}
shows that there are two distinct terms which govern the free evolution of these modes. The first term is the usual Hamiltonian for a massless scalar field in flat space (a collection of harmonic oscillators) whereas the second term is the squeezing interaction, characteristic of the curved space background. Since the background is time-dependent, it sources zero-momentum correlated pairs of the canonical field. This term dominates when $k \ll z'/z \approx aH$, \textit{i.e.} it determines the evolution once the physical wavelengths of these modes become super-Hubble whereas the sub-Hubble modes $k \gg aH$ are in their quantum vacuum, as shown by the first term above. We will assume that the modes are in the Bunch-Davies vacuum. In the presence of these terms in the quadratic Hamiltonian alone, the evolution of the quantum vacuum to the squeezed state is evidently a unitary one \cite{Albrecht:1992kf}.

The story turns interesting once we introduce the cubic non-Gaussianities which imply that there must be higher order interaction terms which lead to mode-coupling between the quantum fluctuations. The cubic action for scalar perturbations contains several terms \cite{Adshead:2011bw,Maldacena:2002vr}, with the leading-order one given by $\mathcal{L}^{(3)} = \int d^3 x\ a\epsilon^2 \zeta (\partial \zeta)^2$.  Going to the interaction picture, operators are written as $\hat{\mathcal{O}}_I = \hat{U}_0^{\dagger} \hat{\mathcal{O}} \hat{U}_0$, where $\hat{U}_0$ is the unitary operator corresponding to the quadractic Hamiltonian \eqref{eq:Hml}. In this formlaism, the interaction Hamiltonian for the Mukhanov-Sasaki variable is given by $\hat{H}_{\rm I}(\tau) = \lambda(\tau) \int d^3 x\,\hat{\chi}(\tau, \vt{x}) (\partial \hat{\chi}(\tau, \vt{x}))^2$, where $\lambda \equiv - a^2 \epsilon^2/z^3$, \textit{i.e.} it is time-dependent. In the interaction picture, time evolution of the density matrix is governed by the von-Neumann equation
\begin{eqnarray}\label{eq:eri}
	\frac{d\rho_I}{d\tau} &=& -i \left[\hat{H}_I (\tau),\rho_I (\tau)\right]\nonumber\\
	& = & - i [\hat{H}_I (\tau), \rho_I (\tau_0)]  - \int_{\tau_0}^{\tau} d\tau' \left\{ \hat{H}_I (\tau) \hat{H}_I (\tau') \rho_I (\tau') - \hat{H}_I(\tau) \rho_I (\tau') \hat{H}_I (\tau') \right. \nonumber \\ 
	&& \left.-
	\hat{H}_I (\tau') \rho_I (\tau') \hat{H}_I(\tau) + \rho_I (\tau') \hat{H}_I (\tau') \hat{H}_I (\tau) \right\}\,,	
\end{eqnarray} 
where, in the second line we have used the formal solution of the density matrix and replaced it back into the equation of motion to bring it to a more amenable form.

In order to describe the cosmological perturbations as part of an open quantum system, we need to write the density matrix in terms of system ($\s$) and environment ($\ev$) dofs. To do so, we define a Hilbert space of quantum fluctuations separated into two regions -- the super- and sub-Hubble mode spaces $\mathcal{H}_\s(t)$ and  $\mathcal{H}_\ev(t)$, respectively. Each of these are tensor products of the Fock space corresponding to the wavenumber $k$, \textit{i.e.} $\mathcal{H}_{\s} = \prod \mathcal{H}_k,\; k < aH$ and similarly for $\mathcal{H}_\ev$, with $k > aH$. We assume that the modes in $\mathcal{H}_{\rm UV}$ (\textit{i.e.} the trans-Planckian modes) can be accounted for by using usual renormalization techniques, and remain agnostic about the quantum gravity theory at play. Note that the important feature is that the boundary between $\mathcal{H}_\s$ and $\mathcal{H}_\ev$ depends on the dynamically expanding background, leading to a time-dependent Hilbert space for the system modes, $\mathcal{H}_\s$, and this is how non-unitarity creeps into the system. 


This splitting also naturally extends for states. For instance, before considering any interactions, the full-state is given by $\ket{\Psi(\tau)} = \ket{SQ(\tau)}_{k<aH} \otimes \ket{0}_{k>aH}$, where $\ket{SQ(\tau)}$ denotes the squeezed state of super-Hubble modes induced by the second term on the r.h.s of Eq. \eqref{eq:Hml}, and $\ket 0$ represents the Bunch-Davies vacuum. Likewise, operators are also split in the same fashion according to the comoving momenta of their Fourier modes. Hence, the interaction Hamiltonian can be expressed as (see the Appendix for more details):
\begin{eqnarray}\label{Eqn:3}
	\hat{H}_I (\tau)  =  \lambda(\tau) \int d^3 x\ \ssm{}{\tau,\vt{x}}  \left(\partial \enm{}{\tau,\vt{x}} \right)^2 =  - \lambda(\tau) \int_{\Delta_k} (\dpr{k_2}{k_3}) \ssm{k_1}{\tau} \enm{k_2}{\tau} \enm{k_3}{\tau}\,,
\end{eqnarray}
where $\int_{\Delta_k} := \int  d^3 k_1 \int d^3 k_2 \int d^3 k_3\ (2\pi)^{-6} \delta (\vt{k_1} + \vt{k_2} + \vt{k_3})$ ensures translation invariance. This particular combination of system and environment modes is the predominant one, since the derivatives favor larger momenta modes (\textit{i.e.} environment modes) and by conservation of momentum, the remaining mode should belong to the system. In the lore of cosmologists, we consider the squeezed limit for the cubic non-Gaussianities. 


Going back to the density matrix, we now assume that at $\tau_0$ the system and environment are not entangled. This, together with a {\it weak coupling} between $\s$ and $\ev$, implies $\rho_I (\tau) = \rho_\s (\tau) \otimes \rho_\ev (\tau_0)$. Then, one can trace out the environmental dofs to find the  {\it reduced density matrix}, $\rho_r (\tau) = \Tr_\ev \left[\rho_I (\tau)\right]$, and its evolution through Eq. \eqref{eq:eri}, such that
	\begin{eqnarray}\label{meq}
		\rho_r'(\tau) & = & \int \frac{d^3 p}{(2\pi)^3}\ \lambda(\tau) \int_{\tau_0}^{\tau} d\tau'\ \lambda(\tau') \left\{\ssm{p}{\tau}\ssm[-]{p}{\tau'} \rho_r (\tau') K_{p}(\tau,\tau') - \ssm{p}{\tau}\rho_r(\tau')\ssm[-]{p}{\tau'} K_{p}^* (\tau,\tau') \right. \nonumber \\ 
		&& - \left. \ssm[-]{p}{\tau'} \rho_r (\tau') \ssm{p}{\tau} K_{p}(\tau,\tau') + \rho_r (\tau') \ssm[-]{p}{\tau'} \ssm{p}{\tau} K_{p}^* (\tau,\tau') \right\}\,,
	\end{eqnarray}
where we have considered the Fourier expansion of the different fields, together with $\rho_\ev(\tau_0) = \ket{0}\bra{0}$, to arrive at the {\it master equation} above (the  {\it Born approximation}). Furthermore, the {\it Born-Markov approximation}, which assumes the temporal locality of the correlations of environment dofs, allows one to write the master equation in terms of Lindblad operators and depict the non-unitary evolution of the density matrix, and are well-suited to study the suppression of coherence \cite{Puri2001, Joos2003}. Since these can be interpreted as randomly varying terms, they are equivalent to the stochastic inflation formalism \cite{Starobinsky:1986fx} and is, thus, a complementary way to study open quantum systems. In general, non-unitarity dynamics are introduced by tracing out environment dofs, such that dissipation-like terms can be extracted from the master equation and its solution \cite{Prokopec:2006fc,Koksma:2010dt}. This behaviour cannot be captured by, and is not equivalent to, calculating loop corrections in an ordinary QFT. Although we shall follow a perturbative method to solve for the master equation, one could have calculated Lindblad operators in a similar way for our cosmological setup \cite{Burgess:2014eoa,Shandera:2017qkg,Nelson:2016kjm}. 

In the master equation above, $K_p (\tau,\tau')$ is typically known as the {\it kernel}, whose solution depends sensitively on the Bunch-Davies mode functions assumed to characterize the initial state. Details of this calculation are shown in the Appendix and the leading-order behaviour, in $\tau$, is given by
\begin{equation}\label{kernel}
	K_{p}(\tau,\tau') \approx - \frac{e^{2i (\tau-\tau')/\tau} \left[1 - e^{-i p (\tau-\tau')}\right] \left[\tau - (1-i)\tau'\right]^2}{8\pi^2 p \tau^4 (\tau')^2 (\tau-\tau')^2}\,.
\end{equation}

In order to solve the master equation \eqref{meq}, we make the first-order approximation that $\rho_r (\tau'') \rightarrow \rho_r (\tau_0)$, which allows for the reduced density matrix to be solved perturbatively (see the Appendix for details).
From the zeroth-order approximation, one gets an almost scale-invariant power spectrum produced by quantum fluctuations, as follows
\begin{align}\label{Eq:6}
	\Delta^2_{\zeta}(q) = \frac{q^3}{2\pi^2 z^2} \left\langle \ssm{\vt{q}}{\tau}\ssm{-\vt{q}}{\tau}\right\rangle = \frac{q^3}{2\pi^2 z^2}  {\rm Tr}\left[\ssm{\vt{q}}{\tau}\ssm{-\vt{q}}{\tau} \rho_r (\tau_0) \right] \approx \frac{1}{2\epsilon\Mp^2} \left(\frac{H}{2\pi}\right)^2\,.
\end{align}

\begin{figure}[t]
\captionsetup[sub]{justification=centering}
  \begin{subfigure}[b]{0.47\textwidth}
    \includegraphics[width=\textwidth]{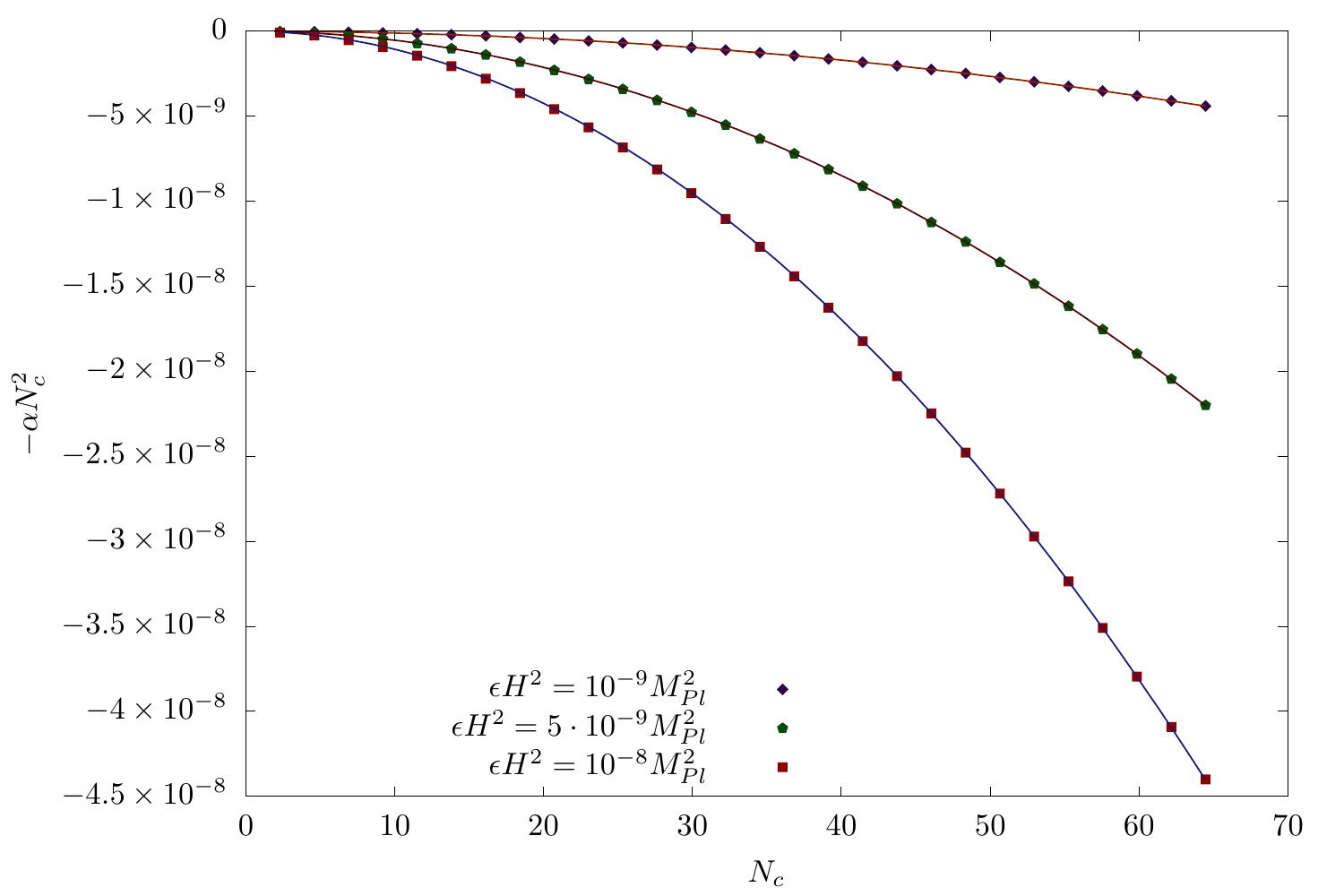}
    \caption{}
  \end{subfigure}
\hfill
  \begin{subfigure}[b]{0.47\textwidth}
    \includegraphics[width=\textwidth]{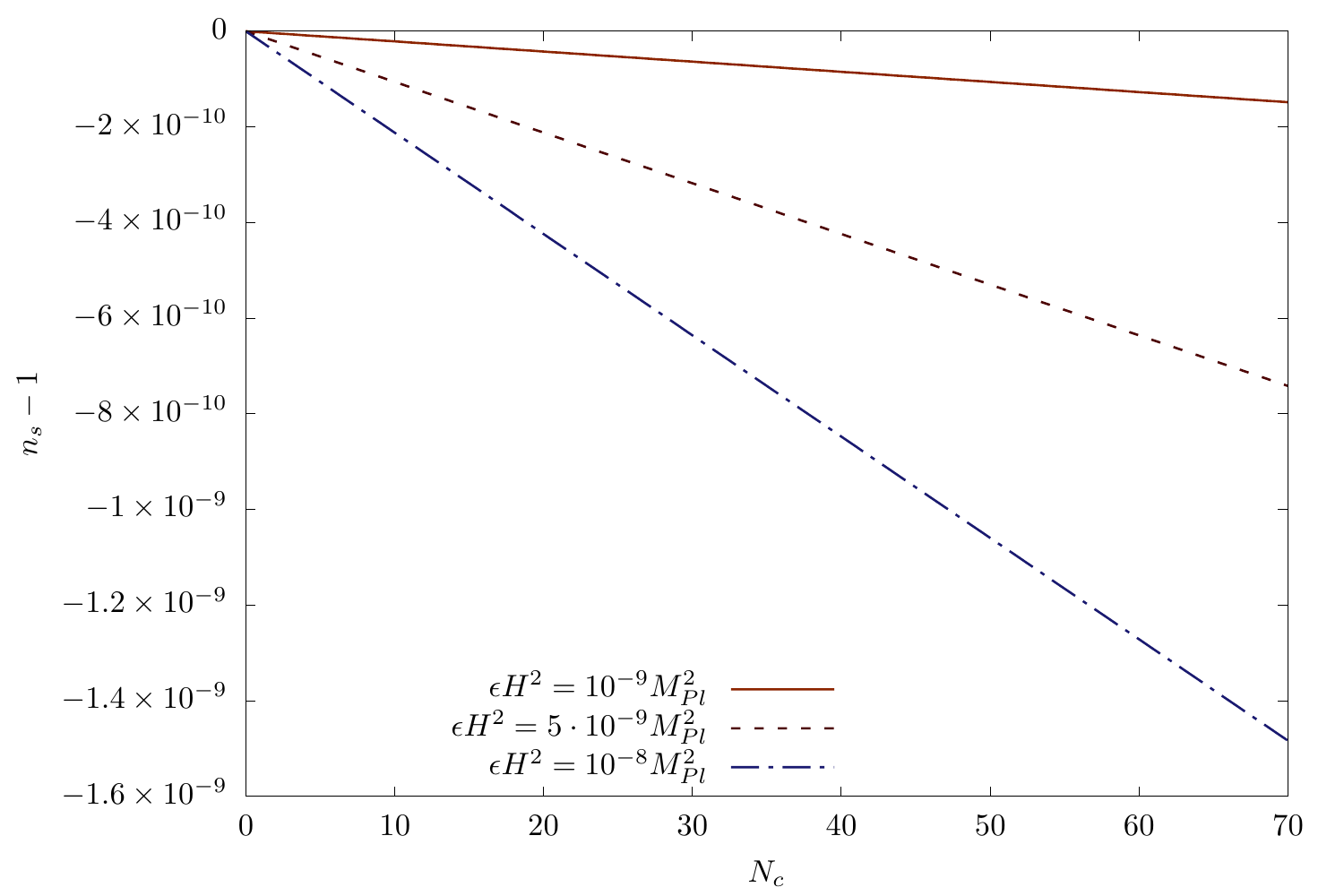}
    \caption{}
  \end{subfigure}
  \caption{Corrections to the power spectrum due to entanglement effects. (a): For each case a quadratic polynomial in $N_c$ fits the data. The free parameter $\alpha$ is found to be determined by the product $\epsilon H^2$. Fixing $\epsilon= 0.01$ and $H^2 \sim M^4_{\rm GUT}/\Mp^2$, consistent with an energy scale of inflation close to the grand unification scale, the correction to the power spectrum is of the order of $\mathcal{O}(10^{-8})$ for a period of $N_c \sim 10^2\ e-$folds of expansion. (b): Deviation from scale invariance due to entanglement between sub-Hubble and super-Hubble modes. The linearity of the functions imply that $\alpha$ corresponds to the running of the spectral index in this range. Corrections to the spectral index, and its running, are of the order $\mathcal{O}(10^{-9})$ and $\mathcal{O}(10^{-11})$, respectively, for the above-mentioned values.}
   \label{Fig2}  
\end{figure}

The first order corrections to the power spectrum come from the perturbative solution of the reduced density matrix Eq. \eqref{meq}, while also substituting the explicit form of the kernel Eq. \eqref{kernel} in the final expression. In this calculation, we encountered multiple integrals, some of which had to be evaluated numerically since an analytical simplification could not be made. Furthermore, several terms in the final answer diverge, as a result of taking the equal time limit in the kernel Eq. \eqref{kernel}, which can be taken care of through renormalization by adding the appropriate counterterms. All of these details have been provided in the Appendix, and we arrive at:
\begin{equation}
	\Delta^2_{\zeta} (q\tau) = \frac{1}{2\epsilon \Mp^2} \left(\frac{H}{2\pi}\right)^2 f\left(N_c\right)\,,
\end{equation}
where $N_c:=\ln(-1/q\tau)$ specifies the number of $e-$folds the Universe has expanded after a Fourier mode crossed the horizon ($q\tau_* = -1$) and became part of the system. The `entanglement function' $f\left(N_c\right) = (1 - \alpha N_c^2)$ has been calculated up to first order in perturbation theory (see Appendix). The quadratic coefficient is given by $\alpha \approx 0.00211886\, \epsilon H^2/(2\Mp^2)$ and is calculated numerically from the integrations carried out in evaluating the two-point function using the reduced density matrix. Notice this formula is valid for any mode, from the very first sub-Hubble mode that crossed the horizon (for which $N_c$ also denotes the total amount of inflation), all the way to the observable modes relevant for the CMB. The first thing to note is that the correction term is suppressed by a factor of $\epsilon H^2/\Mp^2$, which could have been estimated from power counting as loop corrections to the two-point function of curvature perturbations, once everything is normalized and factors of the vertex operators (involving factors of the interaction parameter $\lambda$ of the cubic term under consideration) and propagators are taken into account. The dependence on $N_c^2$ can be intuitively understood as making this correction larger with time -- as entanglement entropy builds with more and more modes turning superhorizon \cite{Brahma:2020zpk}, with the precise numerical prefactor in $\alpha$ resulting from numerical estimations in our open EFT and, thus, shows that our calculation involves dissipative effects over and above straightforward radiative corrections \cite{Senatore:2009cf, Pimentel:2012tw, Seery:2007we} for the power spectrum. This brings up an interesting feature of our result -- to be within the \textit{regime of perturbative validity}, we find an upper bound on the duration of inflation. Since this implies a limit of $N_c \lesssim \alpha^{-1/2}$, using the specific values displayed in Fig. \ref{Fig2}(b), one finds a bound of order $10^5 - 10^6\ e-$folds, which although sufficiently long to solve the standard cosmological puzzles, puts an upper limit coming from perturbativity \cite{Assassi:2012zq}.

On the other hand, such secular divergences in a QFT in quasi-dS backgrounds usually arise from infrared effects \cite{Burgess:2015ajz, Starobinsky:1994bd, Gautier:2013aoa} (see \cite{Urakawa:2010it, Woodard:2015kqa} for some contrarian views). This naturally lead us to consider resummation of the master equation, yielding a nonperturbative result for the entanglement function given by $f(N_c) \simeq e^{-\alpha\,N_c^2}$. An exact analytic calculation using dynamical renormalization group \cite{Burgess:2009bs} is difficult to perform in the absence of exact dS. Nevertheless, one can estimate the resummed function and we sketch the details of the calculation in the Appendix. Other approaches to nonperturbative resummations have also yielded similar results for analogous systems \cite{Boyanovsky:2015tba, Burgess:2009bs}. Interestingly, note that this result still carries signatures of entanglement in the form of the resummed function, as opposed to standard one-loop corrections in which one does not expect any additional time-dependence to appear \cite{Senatore:2009cf, Pimentel:2012tw, Senatore:2012ya}. The nonperturbative result simply ensures that the time-dependence is milder, and not secularly-divergent, as perturbative theory would seem to imply.

Coming to observable consequences, Fig. \ref{Fig2}(a) shows the numerical estimates for the predicted correction to the tree-level value of the power spectrum and the spectral index, for typical values of $\epsilon H^2$. These results are independent of whether we consider the resummed function or simply the first-order correction. Importantly, this is exactly opposite to the predictions for the spectral tilt $(n_s-1)$, and its running, for standard inflationary models \textit{without considering entanglement} \cite{Kosowsky:1995aa}: In that case, both these quantities typically decrease (in magnitude) with increasing amounts of $N_c$ for the modes of interest \cite{Easther:2006tv}. 

One might be tempted to think that the numerical estimates for our correction, as shown in Fig. \ref{Fig2}(a) and Fig. \ref{Fig2}(b), can be made negligibly small by fine-tuning the value of $\epsilon$ to be arbitrarily tiny. However, this is not quite the case since apart from the cubic interaction term considered in this work, there exists another exactly similar term but now with the coefficient $\epsilon \eta$. Given the current value of spectral tilt as measured by the Planck team, it implies that either of $\epsilon$ or $\eta$ must be an $\mathcal{O}(0.01)$ number. It is, therefore, clear that our result would hold even for such fine-tuning. This point has been emphasized in the context of decoherence (and effective stochastic dynamics) \cite{Nelson:2016kjm}, where this effect is caused by the same leading-order terms we consider in this work for calculating the entanglement between the modes. Therefore, our result reinforces the idea that decoherence is not only important for conceptual reasons but also has significance for cosmological observations \cite{Martin:2018zbe}.

The generality of our effect is highlighted by the fact that this should be present for any early-universe scenario which posits that macroscopic inhomogeneities originate from quantum fluctuations. For any such scenario, the horizon will always act as the delimiter between the system and unobservable environment modes and calculations of primordial statistics should be done within an open quantum system. If no such corrections are observed in the future, it would unequivocally state that cosmological perturbations cannot have a quantum origin (but rather have a different mechanism, \textit{e.g.} warm inflation \cite{Berera:1995wh,Berera:1995ie}). Since  dynamics of the background will affect the exact form of the mode functions and also enter the pre-factor of the interaction term, we cannot estimate the magnitude of the corrections in other such scenarios without an explicit computation. Nevertheless we can confidently state their existence. Moreover, it is straightforward to extend our computation to  primordial gravitational waves, the leading order interaction term  for which is independent of slow-roll parameters and therefore, the expected correction to the tensor power spectrum due to entanglement may be larger than the scalar one (in relative magnitude to the tree-level value). These will be pursued in the future and this work forms the foundation of exploring such hitherto undiscovered quantum effects in cosmology.

\bigskip

\noindent {\bf Acknowledgements:}
SB is supported in part by the NSERC (funding reference \#CITA 490888-16) through a CITA National Fellowship and by a McGill Space Institute fellowship. AB is partially supported by STFC. JCF is supported by the Secretary of Higher Education, Science, Technology and Innovation of Ecuador (SENESCYT).

\section*{Appendix}\label{Appendix}
In the appendix, we fill in the details of the derivations for some of the equations used in the main text. In particular, we show detailed derivations of the \textit{kernel}, perturbative solution of the \textit{master equation} and how the corrections to the power spectrum can be derived from this solution, and its nonperturbative resummation.

\section*{System \& Environment Hilbert spaces}
To begin with, let us set the stage by expressing the operators in terms of the system ($\s$) and environment ($\ev$) modes. For this, we filter the sub and super-horizon modes through a window function, where the simplest choice is the Heaviside step function, $\h$. Then, the Fourier expansion of any given operator, like the comoving curvature perturbation, is 
\begin{eqnarray}
	\hat{\zeta} (\tau,\vt{x}) & = & \frac{1}{z}\int \frac{d^3 k}{(2\pi)^3} \left[\chi_k (\tau) \cm{k} + \chi_k^{*} (\tau) \cpp{k} \right] \left[\h(aH-k) + \h(k-aH)\right] e^{i \dpr{k}{x}} \nonumber \\
	& = & \frac{1}{z} \int \frac{d^3 k}{(2\pi)^3}  \left[\left(\chi^{_\s}_k(\tau) + \chi^{_\ev}_k(\tau)\right) \cm{k} + \left(\chi^{_\s}_k(\tau)^* + \chi^{_\ev}_k(\tau)^*\right) \cpp{k} \right] e^{i \dpr{k}{x}} \\
	& = & \hat{\zeta}^{_\s}(\tau, \vt{x})+ \hat{\zeta}^{_\ev}(\tau, \vt{x})\,,
\end{eqnarray}
where
\begin{eqnarray}
	\chi^{_\s}_k(\tau) & = & \chi_k (\tau) \h(aH - k) = \chi_k (\tau) \h(1 + k\tau)\,, \\
	\chi^{_\ev}_k(\tau) & = & \chi_k (\tau) \h(k-aH) = \chi_k (\tau) \h(-1 - k\tau)\,.
\end{eqnarray}
For later convenience, let us also write explicitly the Bunch-Davies functions, the corresponding  state for which we denote by $|0\rangle$:
\begin{equation}\label{BD}
	\chi_k (\tau) = \frac{e^{-i k \tau}}{\sqrt{2k}} \left(1 - \frac{i}{k\tau}\right)\,.
\end{equation}

\section*{Interaction Hamiltonian \& the evolution equation}
Once we have split the fields into operators acting on system and environment states, the interaction Hamiltonian (in the interaction picture) can thus be written as
\begin{equation}
	\hat{H}_I (\tau) = \lambda (\tau) \int d^3 x \left[\enm{}{\tau,\vt{x}} + \ssm{}{\tau,\vt{x}}\right] \left[\partial\left(\enm{}{\tau,\vt{x}} + \ssm{}{\tau,\vt{x}}\right)\right]^2,
\end{equation}
where there are 8 different combinations of system and environment terms (64 considering the Fourier expansion of each field). As explained in the main text, the dominant terms in the integral are associated to the derivatives of environment operators, which can be seen from its Fourier transform. Moreover, since we are ultimately interested in the effect of the environment on the system (and because of momentum conservation), the remaining field left in the cubic term should act on ${\cal H}_\s$. In an alternative approach to that presented in this work, one can work out the Dyson expansion to compute perturbatively the state vectors and then construct the reduced density matrix. In doing so, it is easy to prove that at the same order in perturbation theory, some of the other combinations cancel out. In particular, this happens for $\hat{H}_I (\tau) \sim \enm{}{\tau,\vt{x}}^3$, which is not important as far as corrections to the power spectrum go -- since no system modes are involved -- but it does prove that the kernel does not influence environment modes, or in other words, environment modes are not affected by the (rest of the) environment.

With these considerations, the interaction Hamiltonian we will work with is given by (as shown in Eq. \eqref{Eqn:3}):
\begin{equation}\label{HI}
	\hat{H}_I (\tau)  =  \lambda(\tau) \int d^3 x\ \ssm{}{\tau,\vt{x}}  \left(\partial \enm{}{\tau,\vt{x}} \right)^2  = - \lambda(\tau) \int_{\Delta_k} (\dpr{k_2}{k_3}) \ssm{k_1}{\tau} \enm{k_2}{\tau} \enm{k_3}{\tau}\,,
\end{equation}
where 
$$\int_{\Delta_k} = \int  \frac{d^3 k_1}{(2\pi)^3} \int \frac{d^3 k_2}{(2\pi)^3} \int \frac{d^3 k_3}{(2\pi)^3}\ (2\pi)^3 \delta (\vt{k_1} + \vt{k_2} + \vt{k_3}).$$

Having defined system and environment fields, we shall now focus on the density matrix. In order to make contact with observations one must trace out the high-energy degrees of freedom (dof), which renders a (reduced) density matrix $\rho_r$ acting on ${\cal H}_\s$. This operation is applied to both sides of Eq. \eqref{eq:eri}, yielding the master equation that dictates the evolution of $\rho_r$. However, before doing so, we need to invoke some typical approximations, which are briefly mentioned in the main text.

First, it is assumed that at the beginning of inflation system and environment are not correlated, so the initial state reads $\ket{\Psi (\tau_0)} = \ket{\s_0} \otimes \ket{\ev_0}$, and the density matrix $\rho_I (\tau_0) = \rho_\s (\tau_0) \otimes \rho_\ev (\tau_0)$. A second assumption is that the environment does not change due to its interaction with the system, or in other words, it behaves as a proper environment as is typically required for decoherence. This implies $\rho_\ev (\tau) \approx \rho_\ev (\tau_0)$, which is possible due to the weak coupling between system and environment. Bringing these facts together, the density matrix at later times reads $\rho_I (\tau) = \rho_\s (\tau) \otimes \rho_\ev (\tau_0)$, which facilitates taking the trace over ${\cal H}_\ev$ on Eq. \eqref{eq:eri}. In doing so, we get
\begin{eqnarray}
	\rho'_r (\tau) &=& - \lambda(\tau) \int_{\tau_0}^{\tau} d\tau'\ \lambda(\tau') \int_{\Delta_p} \int_{\Delta_k} (\dpr{p_2}{p_3})(\dpr{k_2}{k_3})  \left\{ \ssm{p_1}{\tau} \ssm{k_1}{\tau'}\times \right.\nn\\
	&& \left.\Tr_\ev \left[\enm{p_2}{\tau}\enm{p_3}{\tau}\ \enm{k_2}{\tau'}\enm{k_3}{\tau'} \rho_\ev (\tau_0) \right] \rho_r (\tau') \right. \nonumber \\
	&& - \ssm{p_1}{\tau} \Tr_\ev \left[\enm{p_2}{\tau}\enm{p_3}{\tau} \rho_\ev (\tau_0) \enm{k_2}{\tau'}\enm{k_3}{\tau'}\right] \rho_r (\tau') \ssm{k_1}{\tau'} \nonumber \\
	&& -  \ssm{k_1}{\tau'} \Tr_\ev \left[\enm{k_2}{\tau'}\enm{k_3}{\tau'} \rho_\ev (\tau_0) \enm{p_2}{\tau}\enm{p_3}{\tau}\right] \rho_r(\tau') \ssm{p_1}{\tau} \nonumber \\
	&& \left. + \rho_r(\tau') \ssm{k_1}{\tau'} \Tr_\ev \left[\rho_\ev (\tau_0) \enm{k_2}{\tau'}\enm{k_3}{\tau'} \enm{p_2}{\tau}\enm{p_3}{\tau}\right] \ssm{p_1}{\tau} \right\}\,,
\end{eqnarray}
where the first order effects arising from $\Tr_\ev ([H_I (\tau), \rho_I (\tau_0)])$ are considered null. To see this as a valid assumption, break the commutator and notice the resulting elements have the form $\sim \ssm{p_1}{\tau} \rho_r (\tau_0) \Tr_\ev \left(\enm{p_2}{\tau}\enm{p_3}{\tau} \ket{0} \bra{0} \right)$, where a non-zero trace requires $\vt{p_2} = - \vt{p_3}$, and by conservation of momentum, $\vt{p_1} = 0$. The contribution from any other combination of momenta is null, and by taking an IR cutoff (which could be given by a small mass induced by the renormalization procedure), the first order terms can be dismissed overall. 

\section*{Master Equation \& the Kernel}
Next, considering the Fourier expansion of the different fields, together with $\rho_\ev(\tau_0) = \ket{0}\bra{0}$, we arrive at the concise form of the master equation, as presented in Eq. \eqref{meq} of the main text.
\begin{eqnarray}
	\rho_r'(\tau) & = & \int \frac{d^3 p}{(2\pi)^3}\ \lambda(\tau) \int_{\tau_0}^{\tau} d\tau'\ \lambda(\tau') \left\{\ssm{p}{\tau}\ssm[-]{p}{\tau'} \rho_r (\tau') K_{p}(\tau,\tau') - \ssm{p}{\tau}\rho_r(\tau')\ssm[-]{p}{\tau'} K_{p}^* (\tau,\tau') \right. \nonumber \\ 
	&& - \left. \ssm[-]{p}{\tau'} \rho_r (\tau') \ssm{p}{\tau} K_{p}(\tau,\tau') + \rho_r (\tau') \ssm[-]{p}{\tau'} \ssm{p}{\tau} K_{p}^* (\tau,\tau') \right\}\,,
\end{eqnarray}
where we have introduced the {\it kernel} as 
\begin{equation}
	K_{p_1} (\tau,\tau') = -2 \int \frac{d^3 p_2}{(2\pi)^3} \left(\dpr{p_2}{p_3}\right)^2 \chi^{_\ev}_{p_2}(\tau) \chi^{_\ev}_{p_2}(\tau')^* \chi^{_\ev}_{p_3}(\tau) \chi^{_\ev}_{p_3}(\tau')^*; \quad \vt{p_3} = - (\vt{p_1} + \vt{p_2})\,.
\end{equation}
In order to compute this integral, we first need to work out the integration limits which are determined by the step functions associated to the fields. For this, let us write the kernel as 
\begin{equation*}
	K_{p_1} \sim \int_{0}^{\infty} d p_2 \int_{-1}^{1} d(\cos \theta) F(p_1,p_2,\theta) \h (p_2 - aH)H_{\theta} (p_2 - (aH)') \h (p_3 - aH) \h (p_3 - (aH)')\,,
\end{equation*}
where $F$ encompasses the product of Bunch-Davies functions Eq. \eqref{BD}, together with the momenta-dependent pre-factors. Further, without loss of generality, we have aligned $\vt{p_1}$ to the $\vt{z}$ direction, so $\theta$ denotes the angle between $\vt{p_2}$ and $\vt{p_1}$ and $p_3^2 = p_1^2 + p_2^2 + 2 p_1 p_2 \cos \theta$.

Next, notice that $\tau' < \tau$ implies $(aH)' < aH$, so the regions in momentum space that contribute to the kernel are solely delimited by the product $\h (p_2 - aH) \h (p_3 - aH)$. Then, the angular region where $\vt{p_3}$ is sub-horizon is delimited by
\begin{equation*}
	p_1^2 + p_2^2 + 2 p_1 p_2 \cos \theta > (aH)^2 \implies \cos \theta > \frac{1}{2 p_1 p_2} \left[ (aH)^2 - (p_1^2 + p_2^2) \right] \equiv \omega\,.
\end{equation*}
This condition sets the lower limit of the integral over the (cosine of the) polar angle, where we also have to account for the possibility $-1 > \omega$ as follows:
\begin{eqnarray*}
	K_{p_1} & \sim & \int_{aH}^{\infty} d p_2 \int_{{\rm max}\{-1, \omega\}}^{1} d(\cos \theta) F(p_1, p_2, \theta) \nonumber \\
	& \sim & \int_{aH}^{\infty} d p_2 \left\{ \int_{-1}^{1} d (\cos \theta) F (p_1,p_2,\theta) \h (-1-\omega) + \int_{\omega}^{1} d(\cos \theta) F(p_1,p_2,\theta) \h (1+\omega) \right\}\,,
\end{eqnarray*}
where
\begin{eqnarray*}
	\h (-1 - \omega) & \implies & -1 \geq \omega \implies p_2 \geq aH + p_1\,, \\
	\h (1 + \omega) & \implies & -1 \leq \omega \implies  p_2 \leq aH + p_1\,.
\end{eqnarray*}
In consequence, the kernel integral is conveniently split as 
\begin{equation}\label{eq:kern2}
	K_{p_1} \sim \int_{aH + p_1}^{\infty} dp_2 \int_{-1}^{1} d(\cos \theta) F(p_1,p_2,\theta) + \int_{aH}^{aH + p_1} dp_2 \int_{\omega}^{1} d(\cos \theta) F(p_1,p_2,\theta)\,.
\end{equation}
In this way, the polar angle $\theta$ and the magnitude of $p_2$ are constrained such that both $p_2$ and $p_3$ are sub-horizon. For small values of $p_1$ -- which already has to satisfy $p_1 < aH$ -- the second integral becomes negligibly small as the lower and upper integration (momentum) limits come closer. 

Performing both integrals, the leading-order behaviour of the kernel is given by (as presented in Eq. \eqref{kernel}):
\begin{equation}
	K_{p_1}(\tau,\tau') \approx - \frac{e^{2i (\tau-\tau')/\tau} \left[1 - e^{-i p_1 (\tau-\tau')}\right] \left[\tau - (1-i)\tau'\right]^2}{8\pi^2 p_1 \tau^4 (\tau')^2 (\tau-\tau')^2}\,.
\end{equation}
The effect from modes in the far-UV (upper limit of the integral) is dismissed by assuming $\tau'$ has a small imaginary part, which is also required for renormalization purposes. This is the same prescription for any interacting field theory, including the {\it in-in} formalism in cosmology, appropriate for closed systems.

\section*{Perturbative solution \& corrections to the power spectrum}
Having derived the explicit expression for the kernel, we can now extract the corrections to the power spectrum through the perturbative solution for the reduced density matrix. For convenience, we rewrite the master equation \eqref{meq}:
\begin{eqnarray}\label{meq1}
	\rho_r'(\tau) & = & \sum_{\vt{p}}\ \lambda(\tau) \int_{\tau_0}^{\tau} d\tau'\ \lambda(\tau') \left\{\ssm{p}{\tau}\ssm[-]{p}{\tau'} \rho_r (\tau') K_{p}(\tau,\tau') - \ssm{p}{\tau}\rho_r(\tau')\ssm[-]{p}{\tau'} K_{p}^* (\tau,\tau') \right. \nonumber \\ 
	&& - \left. \ssm[-]{p}{\tau'} \rho_r (\tau') \ssm{p}{\tau} K_{p}(\tau,\tau') + \rho_r (\tau') \ssm[-]{p}{\tau'} \ssm{p}{\tau} K_{p}^* (\tau,\tau') \right\}\,,
\end{eqnarray}
where for future convenience we have substituted the integral over $p$ for a sum, and a $1/V$ factor is implicit in front of the summation symbol. In order to find a first-order approximation to the solution, we take $\rho_r (\tau'') \rightarrow \rho_r (\tau_0) = \ket{\s_0}\bra{\s_0} = \ket{0}\bra{0}$, which yields
\begin{eqnarray}\label{dmsol}
	\rho_r (\tau) & \approx & \rho_r (\tau_0) + \sum_{\vt{p}} \int_{\tau_0}^{\tau} d\tau' \lambda(\tau') \int_{\tau_0}^{\tau'} d\tau '' \lambda(\tau'') \bigg\{\ssm{p}{\tau'}\ssm[-]{p}{\tau''} \rho_r (\tau_0) K_{p}(\tau',\tau'') \nonumber \\ 
	&& - \ssm{p}{\tau'}\rho_r(\tau_0)\ssm[-]{p}{\tau''} K_{p}^* (\tau',\tau'') - \ssm[-]{p}{\tau''} \rho_r (\tau_0) \ssm{p}{\tau'} K_{p}(\tau',\tau'') \nonumber \\
	&& + \rho_r (\tau_0) \ssm[-]{p}{\tau''} \ssm{p}{\tau'} K_{p}^* (\tau',\tau'') \bigg\}\,.
\end{eqnarray}

Finally, we provide the details for the calculation of the corrections to the power spectrum. In general, it is given by
\begin{equation}
	\Delta^2_{\zeta} (q\tau) = \frac{q^3}{2\pi^2 z^2} \left\langle \ssm{\vt{q}}{\tau}\ssm{-\vt{q}}{\tau}\right\rangle\;,
\end{equation}
where the two-point correlation function is computed through the reduced density matrix technology as follows:
\begin{equation}
	\left\langle \ssm{\vt{q}}{\tau}\ssm{-\vt{q}}{\tau}\right\rangle   =  {\rm Tr}\left[\ssm{\vt{q}}{\tau}\ssm{-\vt{q}}{\tau} \rho_r (\tau) \right]\;.
\end{equation}
In Eq. \eqref{Eq:6}, we showed as a sanity check that the zeroth-order approximation of the reduced density matrix reproduces the well-known scale invariant spectrum. Considering the entanglement effects buried in Eq. \eqref{dmsol}, the correlator is given by
\begin{align}
	\left\langle \ssm{\vt{q}}{\tau}\ssm{-\vt{q}}{\tau}\right\rangle   =&  {\rm Tr}\left[\ssm{\vt{q}}{\tau}\ssm{-\vt{q}}{\tau} \ket{0}\bra{0} \right] + {\rm Tr} \bigg[ \int_{\tau_0}^{\tau} d\tau' \lambda(\tau') \int_{\tau_0}^{\tau'} d\tau'' \lambda(\tau'') \ssm{\vt{q}}{\tau}\ssm{-\vt{q}}{\tau} \nonumber \\
	& \sum_{\vt{p}} \Big\{\ssm{p}{\tau'}\ssm[-]{p}{\tau''} \rho_r (\tau_0) K_{p}(\tau',\tau'') - \ssm{p}{\tau'}\rho_r(\tau_0)\ssm[-]{p}{\tau''} K_{p}^* (\tau',\tau'') \nonumber \\
	& - \ssm[-]{p}{\tau''} \rho_r (\tau_0) \ssm{p}{\tau'} K_{p}(\tau',\tau'') + \rho_r (\tau_0) \ssm[-]{p}{\tau''} \ssm{p}{\tau'} K_{p}^* (\tau',\tau'') \Big\} \bigg]\;,
\end{align}
where the corrections are clearly dictated by the second trace. Then, focusing on this bit, we take out of the sum the terms with momentum $p=q$, rendering
\begin{align}
	{\rm Tr}&\bigg[\ssm{\vt{q}}{\tau}\ssm{-\vt{q}}{\tau} \Big\{\ssm{q}{\tau'}\ssm[-]{q}{\tau''} \ket{0}\bra{0} K_{q}(\tau',\tau'') - \ssm{q}{\tau'}\ket{0}\bra{0}\ssm[-]{q}{\tau''} K_{q}^* (\tau',\tau'') \nonumber \\
	&- \ssm[-]{q}{\tau''} \ket{0}\bra{0} \ssm{q}{\tau'} K_{q}(\tau',\tau'') 
	+ \ket{0}\bra{0} \ssm[-]{q}{\tau''} \ssm{q}{\tau'} K_{q}^* (\tau',\tau'') \nonumber \\
	&+ \ssm[-]{q}{\tau'}\ssm{q}{\tau''} \ket{0}\bra{0} K_{q}(\tau',\tau'') - \ssm[-]{q}{\tau'}\ket{0}\bra{0}\ssm{q}{\tau''} K_{q}^* (\tau',\tau'') \nonumber \\
	&-\ssm{q}{\tau''} \ket{0}\bra{0} \ssm[-]{q}{\tau'} K_{q}(\tau',\tau'') 
	+ \ket{0}\bra{0} \ssm{q}{\tau''} \ssm[-]{q}{\tau'} K_{q}^* (\tau',\tau'')  + \sum_{\vt{p} \neq \pm \vt{q}} \ldots
	\Big\} \bigg]\;.
\end{align}
It is easy to show that the last sum vanishes, whereas the other terms, which involve momenta with magnitude $q$, lead to
\begin{align}\label{cps0}
	\left\langle \ssm{\vt{q}}{\tau}\ssm{-\vt{q}}{\tau}\right\rangle = & \frac{1}{2q} \left(1+\frac{1}{(q\tau)^2}\right) +   2 \int_{-1/q}^{\tau} d\tau' \lambda(\tau') \int_{-1/q}^{\tau'} d\tau'' \lambda(\tau'') \nonumber \\
	& \Big\{K_q (\tau', \tau'') \left[\left(\chi_q(\tau)\right)^2 \chi_q^*(\tau') \chi_q^*(\tau'') - \left|\chi_q(\tau) \right|^2 \chi_q (\tau')\chi_q^* (\tau'') \right] + {\rm c.c.} \Big\}\;,
\end{align}
where the lower limits of the integrals ``change'' due to the step functions associated to the system operators. These limits imply that the environment will influence a particular mode only after it crosses the horizon, as one would expect. 

Introducing the change of variable $w = q \tau$, the scalar power spectrum including entanglement effects can be written as
\begin{align}\label{cps}
	\Delta^2_{\zeta} (w) = & \frac{1}{2\epsilon \Mp^2} \left(\frac{H}{2\pi}\right)^2 \bigg[ 1 + w^2 + \frac{\epsilon H^2}{2 \Mp^2} w^2 \int_{-1}^{w} dw' w' \int_{-1}^{w'} dw'' w'' \nonumber \\
	& \Big\{K (w', w'') \left[\left(\chi(w)\right)^2 \chi^*(w') \chi^*(w'') - \left|\chi(w) \right|^2 \chi (w')\chi^* (w'') \right] + {\rm c.c.} \Big\} \bigg]\;,
\end{align}
where any $q-$dependence on the functions is either absorbed by the new variable or by the momenta in the power spectrum formula. As a point of detail, notice that the (isolated) factor of $w^2$ corresponds to the decaying mode, which was seen to be subdominant with respect to every other term (after a few e-folds after horizon crossing). 

The integration over $w''$ can be done analytically, but the subsequent integration over $w'$ has to be performed numerically. In doing so, and after removing equal-time divergences (to be discussed soon after), one finds that 
\begin{equation}\label{ps_prt}
	\Delta^2_{\zeta} (w) = \frac{1}{2\epsilon \Mp^2} \left(\frac{H}{2\pi}\right)^2 (1 - \alpha N_c^2)\;,
\end{equation}
where $N_c = -\ln(-w)$ denotes the number of e-folds after horizon crossing, and the constant is given by (decaying mode not considered)
\begin{eqnarray}\label{aph}
	\alpha \approx 0.00211886\frac{\epsilon H^2}{2\Mp^2}\;.
\end{eqnarray}

\section*{Nonperturbative Resummation}

The validity of Eq. \eqref{ps_prt} relies on the validity of perturbation theory. Here, we will present an attempt to resum this result following \cite{Boyanovsky:2015tba}. For this, we will have to make more assumptions, so that, even though we are not dealing with exactly the same case, the following procedure presents a strong evidence that the full power spectrum can be indeed resummed. 

First, the mode functions are split into a growing and a decaying mode,
\begin{equation}
	\ssm{\vt{q}}{\tau} = \chi^{+}_q (\tau)  \hat{P}_{\vt q} - \chi^{-}_q(\tau) \hat{X}_{\vt q} \,,
\end{equation}
where $\chi^{+(-)}_q$ and $\hat{P}_{\vt q} (\hat{X}_{\vt q})$ denote the growing (decaying) mode function and operator respectively, and are given by
\begin{align}
	\chi^{+}_q (\tau) = \sqrt{\frac{-\pi \tau}{2}} Y_{3/2} (|q \tau|)\,, & \qquad \chi^{-}_q (\tau) = \sqrt{\frac{-\pi \tau}{2}} J_{3/2} (|q \tau|)\,, \\
	\hat{P}_{\vt q} = \frac{i}{\sqrt{2}} \left( \cpp{q} - \cm{q} \right)\,, & \qquad \hat{X}_{\vt q} = \frac{1}{\sqrt{2}} \left(\cm{q} + \cpp{q} \right)\,.
\end{align}
The operators satisfy the usual commutation relation $\left[\hat{X}_{\vt q}, \hat{P}_{\vt k}\right] = i \delta_{{\vt q},-{\vt k}}$, with vanishing commutators for equal operators. 

Then, at leading order in $q\tau$, the power spectrum of the growing mode operator captures the behaviour of the full power spectrum, since
\begin{equation}
	\left\langle \ssm{\vt{q}}{\tau} \ssm{-\vt{q}}{\tau} \right\rangle \approx \left[ \chi^{+}_q (\tau) \right]^2 \left\langle \hat{P}_{\vt q}  \hat{P}_{-\vt q} \right\rangle \approx \frac{1}{2q^2 \tau^3}\,,
\end{equation}
in agreement with the leading order term of Eq. \eqref{cps0}. Then, for our purposes it proves easier to deal with this magnitude as opposed to the full power spectrum, which presents an `extra' time dependence. 

Next, we shall re-write the master equation in terms of the growing and decaying mode operators. However, before doing so, it is useful to take the Markov approximation, for which under the assumption of weak coupling one can take $\rho_r (\tau') \rightarrow \rho_r (\tau)$, rendering 
\begin{eqnarray}\label{meq3}
	\rho_r'(\tau) & = & \sum_{\vt{p}}\ \lambda(\tau) \int_{\tau_0}^{\tau} d\tau'\ \lambda(\tau') \left\{\ssm{p}{\tau}\ssm[-]{p}{\tau'} \rho_r (\tau) K_{p}(\tau,\tau') - \ssm{p}{\tau}\rho_r(\tau)\ssm[-]{p}{\tau'} K_{p}^* (\tau,\tau') \right. \nonumber \\ 
	&& - \left. \ssm[-]{p}{\tau'} \rho_r (\tau) \ssm{p}{\tau} K_{p}(\tau,\tau') + \rho_r (\tau) \ssm[-]{p}{\tau'} \ssm{p}{\tau} K_{p}^* (\tau,\tau') \right\}\,.
\end{eqnarray}
In this way one can compute the integrals over $\tau'$ which only include the Bunch-Davies functions and the kernel (or its conjugate). In performing the integrations one finds equal-time divergences (see next section), but for now we shall focus on the late time leading-order behaviour of the integrals which we will exploit in order to cast the master equation in a more convenient way. The integrals are
\begin{align}
	{\cal Y}_q^+ (\tau) & = \int_{\tau_0}^{\tau} d\tau' \lambda(\tau') \chi^{+}_q (\tau') K_q (\tau,\tau') \sim - \int_{\tau_0}^{\tau} d\tau' \lambda(\tau') \chi_q^{+} (\tau') K_q^* (\tau,\tau')  \sim \frac{\epsilon^{1/2} H}{2^{3/2} \Mp} \frac{i}{8\pi^2} \frac{\ln (|q \tau|)}{q^{3/2} \tau^4}\,, \\
	{\cal Y}_q^- (\tau) & = \int_{\tau_0}^{\tau} d\tau' \lambda(\tau')\chi^{-}_q (\tau') K_q (\tau,\tau') \sim - \int_{\tau_0}^{\tau} d\tau' \lambda(\tau') \chi_q^{-} (\tau') K_q^* (\tau,\tau')  \sim \frac{\epsilon^{1/2} H}{2^{3/2} \Mp} \frac{i}{24\pi^2} \frac{ q^{5/2}\ln (|q \tau|)}{\tau} \,,
\end{align}
where we have omitted a subdominant real part, which would become imaginary in the power spectrum, leading to an oscillatory term that is of no interest in the present discussion. With these considerations, the master equations can be written as
\begin{align}
	\rho_r'(\tau)  = \lambda(\tau) \sum_{\vt p} & \bigg\{ \chi_p^+ (\tau) {\cal Y}_p^+ (\tau) \left[ \hat{P}_{\vt p} \hat{P}_{-\vt p} \rho_r (\tau) - \rho_r (\tau) \hat{P}_{\vt p} \hat{P}_{-\vt p} \right]\nn\\
	&  + \chi^-_p (\tau) {\cal Y}_p^- (\tau) \left[ \hat{X}_{\vt p} \hat{X}_{-\vt p} \rho_r (\tau) -  \rho_r (\tau) \hat{X}_{\vt p} \hat{X}_{-\vt p}	\right] \nonumber \\
	& - \chi_p^+ (\tau) {\cal Y}_p^- (\tau) \left[ \hat{P}_{\vt p} \hat{X}_{-\vt p} \rho_r (\tau) - \hat{X}_{-\vt p} \rho_r (\tau) \hat{P}_{\vt p} - \rho_r(\tau) \hat{X}_{-\vt p} \hat{P}_{\vt p} + \hat{P}_{\vt p} \rho_r (\tau) \hat{X}_{-\vt p}  \right] \nonumber \\
	& - {\cal Y}^+_p (\tau) \chi_p^- (\tau) \left[  \hat{X}_{\vt p}  \hat{P}_{-\vt p} \rho_r (\tau) -  \hat{P}_{-\vt p} \rho_r(\tau)  \hat{X}_{\vt p} - \rho_r (\tau)  \hat{P}_{-\vt p}  \hat{X}_{\vt p}  +  \hat{X}_{\vt p} \rho_r (\tau)  \hat{P}_{-\vt p} \right] \bigg\}\,.
\end{align}

Then, one can compute the corrections to this power spectrum, taking advantage of the following equality
\begin{equation}
	\frac{d}{d\tau} \left\langle \hat{P}_{\vt q}  \hat{P}_{-\vt q} \right\rangle = \Tr \left[  \hat{P}_{\vt q}  \hat{P}_{-\vt q} \rho_r' (\tau) \right]\,.
\end{equation}
The trace can be computed by using the commutation relations of $\hat{X}$ and $\hat{P}$, together with the properties of the trace. In doing so, we find that 
\begin{align}
	\Tr \left[  \hat{P}_{\vt q}  \hat{P}_{-\vt q} \rho_r' (\tau) \right] & \approx 4i \lambda (\tau) {\cal Y}^+_q (\tau) \chi^-_q (\tau) \Tr\left[\hat{P}_{\vt q}  \hat{P}_{-\vt q} \rho_r (\tau) \right]\,,
\end{align}
where we are ignoring a subdominant contribution from the product of two decaying modes (the others are null). Hence, we have that
\begin{equation}
	\frac{d}{d\tau} \left\langle \hat{P}_{\vt q}  \hat{P}_{-\vt q} \right\rangle = - \frac{\epsilon H^2}{48\pi^2 \Mp^2} \frac{\ln(|q \tau|)}{\tau} \left\langle \hat{P}_{\vt q}  \hat{P}_{-\vt q} \right\rangle\,.
\end{equation}
Performing the integration, one arrives to 
\begin{equation}
	\left\langle \hat{P}_{\vt q}  \hat{P}_{-\vt q} \right\rangle = f(\tau) \left\langle \hat{P}_{\vt q}  \hat{P}_{-\vt q} \right\rangle (\tau_*)\,,
\end{equation}
where 
\begin{equation}
	f(\tau) = \exp\left[-\frac{\epsilon H^2}{48\pi^2 \Mp^2}  \int_{-1/q}^{\tau} d\tau' \frac{\ln (-q\tau)}{\tau} \right] = \exp\left[-\frac{\epsilon H^2}{96\pi^2 \Mp^2} \ln^2 (|q \tau|)\right]\,.
\end{equation}
Notice that the pre-factor of the logarithm above can be written as 
\begin{equation} \label{aph2}
	\alpha = 0.00211086 \frac{\epsilon H^2}{2\Mp^2}\,,
\end{equation}
which is remarkably close to the parameter $\alpha$ obtained through the perturbative expansion, as shown in Eq. \eqref{aph}. This shows that long after horizon crossing there are other terms that take over the secular term, rendering a well-behaved power spectrum, whose final form is  
\begin{equation}
	\Delta^2_{\zeta} = \frac{1}{2\epsilon \Mp^2} \left(\frac{H}{2\pi}\right)^2 e^{-\alpha N_c^2}\,,
\end{equation}
where $\alpha$ has been estimated numerically (Eq. \eqref{aph}) and analytically (Eq. \eqref{aph2}).

Let us highlight the approximations we had to make in order to arrive at this result. Firstly, as opposed to \cite{Boyanovsky:2015tba}, we are dealing with the long and short wavelength modes of the curvature perturbations themselves. This implies working with mode functions which are much more complicated, and therefore, one has to resort to numerical estimations for the integrals involved while working in the perturbative approach. More importantly, recall that our secular divergence is for an open EFT in quasi-dS background. This requires calculating kernels, as opposed to loop corrections for scalar fields on dS, and thus, an application of the dynamical renormalization group is not entirely straightforward \cite{Burgess:2009bs}. However, interestingly, we find that the resummed `entanglement function' takes a form that is remarkable similar to what one would expect from the dynamical renormalization group analyses \cite{Burgess:2009bs}. In conclusion, even though we arrive at the late-time value of the resummed function within the Markovian approximation, its final form reassures us about its validity.

\section*{Renormalization of equal-time divergences}
Our final missing piece in the calculations comes in the form of the renormalization schemes used in curing the divergences appearing in the master equation. These are ones coming from taking the equal time limit in the kernel, which must be the case, while evaluating the necessary integrals involved in calculating the corrections to the power spectrum. The most appropriate way to deal with these divergences is to determine the counterterms which have to be introduced in order to take care of them. Let us begin with the master equation once again
\begin{eqnarray}\label{meq2}
	\rho_r'(\tau) & = & \sum_{\vt{p}}\ \lambda(\tau) \int_{\tau_0}^{\tau} d\tau'\ \lambda(\tau') \left\{\ssm{p}{\tau}\ssm[-]{p}{\tau'} \rho_r (\tau') K_{p}(\tau,\tau') - \ssm{p}{\tau}\rho_r(\tau')\ssm[-]{p}{\tau'} K_{p}^* (\tau,\tau') \right. \nonumber \\ 
	&& - \left. \ssm[-]{p}{\tau'} \rho_r (\tau') \ssm{p}{\tau} K_{p}(\tau,\tau') + \rho_r (\tau') \ssm[-]{p}{\tau'} \ssm{p}{\tau} K_{p}^* (\tau,\tau') \right\}\;.
\end{eqnarray}
The perturbative expansion instructs $\rho_r (\tau') \rightarrow \rho_r (\tau_0)$, so that applying the operators at $\tau'$ renders
\begin{align}\label{meq3}
	\rho_r'(\tau) = & \sum_{\vt{p}}\ \lambda(\tau) \int_{\tau_0}^{\tau} d\tau'\ \lambda(\tau') \left\{ K_{p}(\tau,\tau') \ssm{p}{\tau}\chi_p^{_\s} (\tau')^* \ket{1_{\vt{p}}}\bra{0}  - K_{p}^* (\tau,\tau') \ssm{p}{\tau}\ket{0}\bra{1_{-\vt{p}}}\chi_p^{_\s} (\tau')  \right. \nonumber \\ 
	& \left. -K_{p}(\tau,\tau') \chi_p^{_\s} (\tau')^* \ket{1_{\vt{p}}}\bra{0} \ssm{p}{\tau}  + K_{p}^* (\tau,\tau') \ket{0}\bra{1_{-\vt{p}}} \chi_p^{_\s} (\tau') \ssm{p}{\tau}  \right\}\;.
\end{align}
Then, we may perform the integrals over $\tau'$, where we will only pay attention to the upper limit, which is the one leading to the {\it equal-time} divergences. The integrals we need to compute are
\begin{eqnarray}
	\int_{\tau_0}^{\tau} d\tau' \tau' K_p (\tau,\tau') \chi_p^{_\s} (\tau')^*\;, \qquad \int_{\tau_0}^{\tau} d\tau' \tau' K_p^* (\tau,\tau') \chi_p^{_\s} (\tau')\;,
\end{eqnarray}
which can be done analytically. In order to isolate the divergence, we take the upper limit as $\tau' \rightarrow \tau - i \delta$, with $0 < \delta \ll 1$, to then Taylor expand the resulting expression around $\delta = 0$. This yields
\begin{align}
	&\int_{\tau_0}^{\tau-i\delta} d\tau' \tau' K_p (\tau,\tau') \chi_p^{_\s} (\tau')^* \sim -\frac{i}{8\pi^2} \frac{\chi_{p}^{_\s}(\tau)^*}{\tau^3} \ln \left(\frac{\delta}{\mu}\right) + {\cal O}(\delta^0)\;, \\
	&\int_{\tau_0}^{\tau-i\delta} d\tau' \tau' K_p^* (\tau,\tau') \chi_p^{_\s} (\tau') \sim \frac{i}{8\pi^2} \frac{\chi_{p}^{_\s}(\tau)}{\tau^3} \ln \left(\frac{\delta}{\mu}\right) + {\cal O}(\delta^0)\;,
\end{align}
where $\mu$ is a renormalization scale. Next, plugging these expressions back to \eqref{meq3} and recovering the full operators, we arrive to 
\begin{align}
	\rho_r'(\tau) = -\frac{i}{8\pi^2} \frac{\epsilon H^2}{8 \Mp^2 \tau^2} \ln \left(\frac{\delta}{\mu}\right) & \sum_{\vt{p}}  \left\{ \ssm{p}{\tau}\ssm[-]{p}{\tau} \rho_r(\tau_0) + \ssm{p}{\tau} \rho_r (\tau_0) \ssm[-]{p}{\tau} \right. \nonumber \\
	& \left. - \ssm[-]{p}{\tau} \rho_r (\tau_0) \ssm{p}{\tau} - \rho_r (\tau_0) \ssm[-]{p}{\tau}\ssm{p}{\tau} \right\} + {\cal O} (\delta^0)\;.
\end{align}
Notice this can be further simplified by taking the second and fourth terms on the r.h.s. with opposite momenta (taking advantage of the sum), so the second and third terms cancel out such that
\begin{equation}
	\rho_r'(\tau) =-\frac{i}{8\pi^2} \frac{\epsilon H^2}{8 \Mp^2 \tau^2} \ln \left(\frac{\delta}{\mu}\right) \sum_{\vt{p}} \left[\ssm{p}{\tau} \ssm[-]{p}{\tau},\rho_r(\tau_0)\right] + {\cal O}(\delta^0)\;.
\end{equation}
Compare this to the first order approximation of the Liouville--von Neumann equation,
\begin{equation}
	\rho'_I (\tau) \approx -i \left[H_I (\tau), \rho_I (\tau_0)\right]\;,
\end{equation}
which suggests the need of a mass counterterm,
\begin{equation}
	\delta {\cal L} = - \delta H = \frac{\delta m^2}{2} \sum_{\vt{p}} \hat{\chi}_{\vt{p}} (\tau) \hat{\chi}_{-\vt{p}}(\tau)\;,
\end{equation}
with
\begin{equation}
	\delta m^2 = \frac{1}{8\pi^2} \frac{\epsilon H^2}{4 \Mp^2 \tau^2} \ln \left(\frac{\delta}{\mu}\right)\;.
\end{equation}
By introducing this counterterm we may just ignore any divergences found in the computation of the correction of the power spectrum found in Eq. \eqref{cps}. The other divergences encountered can be taken care of by similar counterterms and shows that these divergences are spurious and only the finite term arising from this calculation contributes to the power spectrum, as has been shown in the main text. Finally, notice that even though there are no finite contributions at first order, there are infinite terms like those emerging from the IR or the computed counterterm.

\end{document}